\RequirePackage[switch]{lineno}
\documentclass[a4paper]{article}

\usepackage{icrc2013}
\usepackage{xspace}
\usepackage[english]{babel}
\usepackage{amsmath}
\usepackage{amssymb}
\usepackage{color}

\newcommand{\cgcg}{CGCG~291$-$028}
\newcommand{\chandra}{\textit{Chandra}}
\newcommand{\swift}{\textit{Swift}}
\newcommand{\gsi}{\,\raisebox{-0.13cm}{$\stackrel{\textstyle>}
{\textstyle\sim}$}\,}
\newcommand{\be}{\begin{equation}}
\newcommand{\ee}{\end{equation}}

\title{Tentative identification of the source of four UHECRs\\ and implications thereof}

\shorttitle{The source of four UHECRs and implications}

\authors{Glennys R. Farrar}

\afiliations{
Center for Cosmology and Particle Physics, Department of Physics, New York University, NY, NY 10003, USA }

\email{gf25@nyu.edu}

\abstract{Four UHECRs in the combined HiRes and AGASA datasets are backtracked in the Galactic magnetic field.  They point to a common source which is localized to within 1 degree, if they are protons as is shown to be the most probable charge assignment.  A Swift-BAT hard X-ray AGN in the galaxy CGCG 291-028 is the only notable source candidate within the source locus and within the GZK distance horizon.  The spectrum of the four events is consistent with production in a transient event such as a stellar tidal disruption flare.  Under the assumption the UHECRs were produced in CGCG 291-028, the total energy of UHECRs produced by the transient can be estimated and extragalactic magnetic deflections can be constrained.  If CGCG 291-028 is indeed the source of the UHECRs, observations of its present state should elucidate the phenomenon of UHECR acceleration.}
\begin{document}
\maketitle

\section{The Ursa Major UHECR cluster}

\begin{figure}
\centering
\vspace{-0.5 in}
\includegraphics[width=0.5\textwidth]{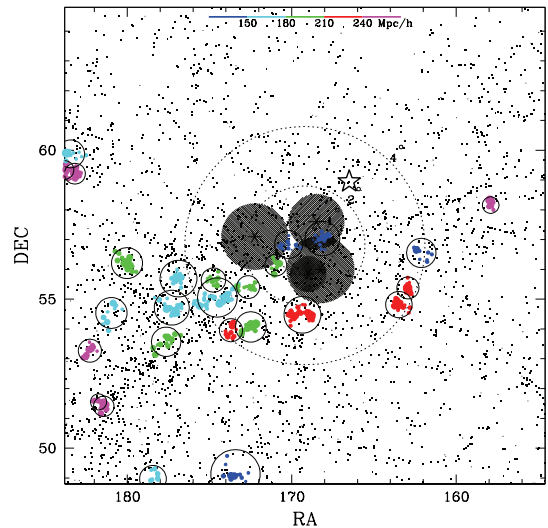}
\vspace{-1.1 in}
\caption{\label{uhecr_image} UMC arrival directions with 68\% uncertainty domains and SDSS galaxies \cite{fbh06}, in equatorial coordinates.}
\vspace{-0.4 in}
\end{figure}

An analysis of the AGASA and binocular HiRes UHECR datasets revealed \cite{HRGF} four events -- three from AGASA and one from HiRes -- whose arrival directions are much closer than is likely by chance\cite{f07}; these UHECRs are designated the ``Ursa Major'' cluster (UMC) below.  Correcting the experiments' absolute energy calibrations by the factors determined by the Auger-TA working group \cite{EnergyCorrections}, 0.65 and 0.91 respectively, and using the final HiRes event reconstruction\cite{HRfinal}, gives the events' arrival directions and energies shown in Table \ref{UMevts}. Fig. \ref{uhecr_image} shows the observed arrival directions of the UMC UHECRs in Equatorial Coordinates with their positional uncertainties, along with galaxies from SDSS (clusters shown in color) \cite{fbh06}.  With 97 events in the combined HiRes+AGASA dataset above (energy-renormalized) 30 EeV, and a field of view of approximately 20,000 sq deg, the probability of finding 4 events in a single $\approx$ 10 sq-degree area by chance is $\approx 0.2$\%.  

There is a huge void in the large scale matter distribution in the direction of the UMC, extending to nearly 200 Mpc\cite{fbh06}, so that extragalactic deflections can be expected to be exceptionally small.  Furthermore, thanks to recent progress in determining the Galactic magnetic field (GMF), the events can be backtracked to determine their source direction;  if they are protons their arrival directions are consistent with having been produced in a single source, as discussed in Sec. \ref{magdefs}.  Their spectrum suggests that the UHECRs were produced in a transient event (Sec. \ref{burst}), and in Sec. \ref{src} a tentative identification of the host galaxy is proposed.  Sec. \ref{interp} discusses some of the many interesting inferences which follow, if future observations confirm this interpretation.

\begin{table}
\centering
\caption{\label{UMevts} The energies and arrival directions (J2000 Galactic coordinates) of the 3 AGASA and 1 HiRes events in the UM cluster, with radius of the angular region containing 68\% of the arrival probability in the last column.}
\vspace{0.1 in}
\begin{tabular}{|c|c|c|c|}
  \hline
  Energy (EeV) & $L $ & $B$ & $\sigma_{{68}} $\\
  \hline
50.5 & 145.5$^{\circ}$ & 55.1$^{\circ}$ & 1.2$^{\circ}$        \\
 35.7 & 143.2$^{\circ}$ & 56.6$^{\circ}$ & 1.4$^{\circ}$        \\
  34.8 & 147.5$^{\circ}$ & 56.2$^{\circ}$ & 1.4$^{\circ}$        \\
   36.6 & 143.6$^{\circ}$ & 58.4$^{\circ}$ & 0.8$^{\circ}$        \\
    \hline
\end{tabular}
\vspace{-0.1 in}
\end{table}

\vspace{-0.2 in} 
\section{Galactic magnetic field, UHECR deflection mapping and locus of UM source}\label{magdefs}

Jansson and Farrar\cite{JF12,JF12rand} (JF12 below) have developed a 35-parameter model of the Galactic magnetic field composed of i) a coherent, large-scale regular field, with disk, halo and poloidal components, ii) a striated (ordered) random field, and iii) a purely random field.  The model parameters are constrained with 37391 Rotation Measures (RMs) of extragalactic sources, and the WMAP polarized and unpolarized synchrotron emission (Q, U and I) maps.  These are smoothed to 13.4-square-degree pixels and for each of these pixels, the uncertainty on the mean value is determined by the variance of the data within in, which is due to astrophysical effects such as inhomogeneities.  The parameters of the model are fit by minimizing the chi-squared between predicted and observed RM, Q,U, and I maps.  The resultant model gives a good overall description of the data, with $\chi^2$ per degree of freedom of 1.096 for the fit of the regular and striated fields using the RM and polarized synchrotron observables (6605 degrees of freedom) and 1.064 for the fit to the random field using the total synchrotron emission (2957 degrees of freedom).  This description is dramatically better than obtained with earlier field models which did not allow for poloidal and striated components\cite{JF12rand}.  

The deflections of UHECRs in the JF12 coherent GMF have now been determined over the entire sky, for rigidity ($R \equiv E/Z$) above 1 EeV, with angular resolution $\approx 0.1^{\circ}$\cite{fjr13}.  To do this, CRs are isotropically backtracked using NASA's Pleiades supercomputer, covering the sky with one CR in each Healpix res-11 pixel, about 50M pixels, at each of 20 steps of log($R) = 0.05$.  Liouville's theorem implies that in a uniformly illuminated Galaxy, the illumination at Earth is also uniform.  Therefore, this procedure yields the magnification value for each extragalactic source direction, in the JF12 regular GMF.  Using database technology as described in \cite{fjr13}, the backtracking information is efficiently converted to forward-tracking maps at each of the 20 rigidities.  A web-based tool to do this has been developed by J. Roberts, NYU, with the expectation of making it publically available in due course; it is described in \cite{fjr13}.  The deflection analysis is currently being extended to include the random and striated components of the JF12 magnetic field\cite{fjkrs13}.

For most UHECRs, deflection by the GMF produces a large uncertainty in the source direction, but this is not the case for the UMC events for two reasons.  First, the magnetic deflections along this line of sight are small for rigidities $\geq 35$ EeV, as is the case if the UMC events are protons \cite{fjr13}; this limits the deflection uncertainties from both the GMF itself and from energy resolution.  Secondly, although the charge of a UHECR is not generally known, which ordinarily produces a large uncertainty as to its deflection, the UMC events can be inferred to most likely all have $Z=1$, as follows: i) The void in the foreground\cite{fbh06} implies a minimum propagation distance too great for intermediate mass nuclei to survive, limiting the possible charge assignments to $Z=1$ or heavy ($Z > 20$).  ii) With $Z=1$, all events backtrack to a common locus, whereas for large $Z $, i.e., low rigidity, the deflections are very large, so the cluster would have to be a chance association, but this has only a $\approx 0.2$\% probability as noted above.  

\begin{figure}[t!]
\centering
\includegraphics[width=0.4\textwidth]{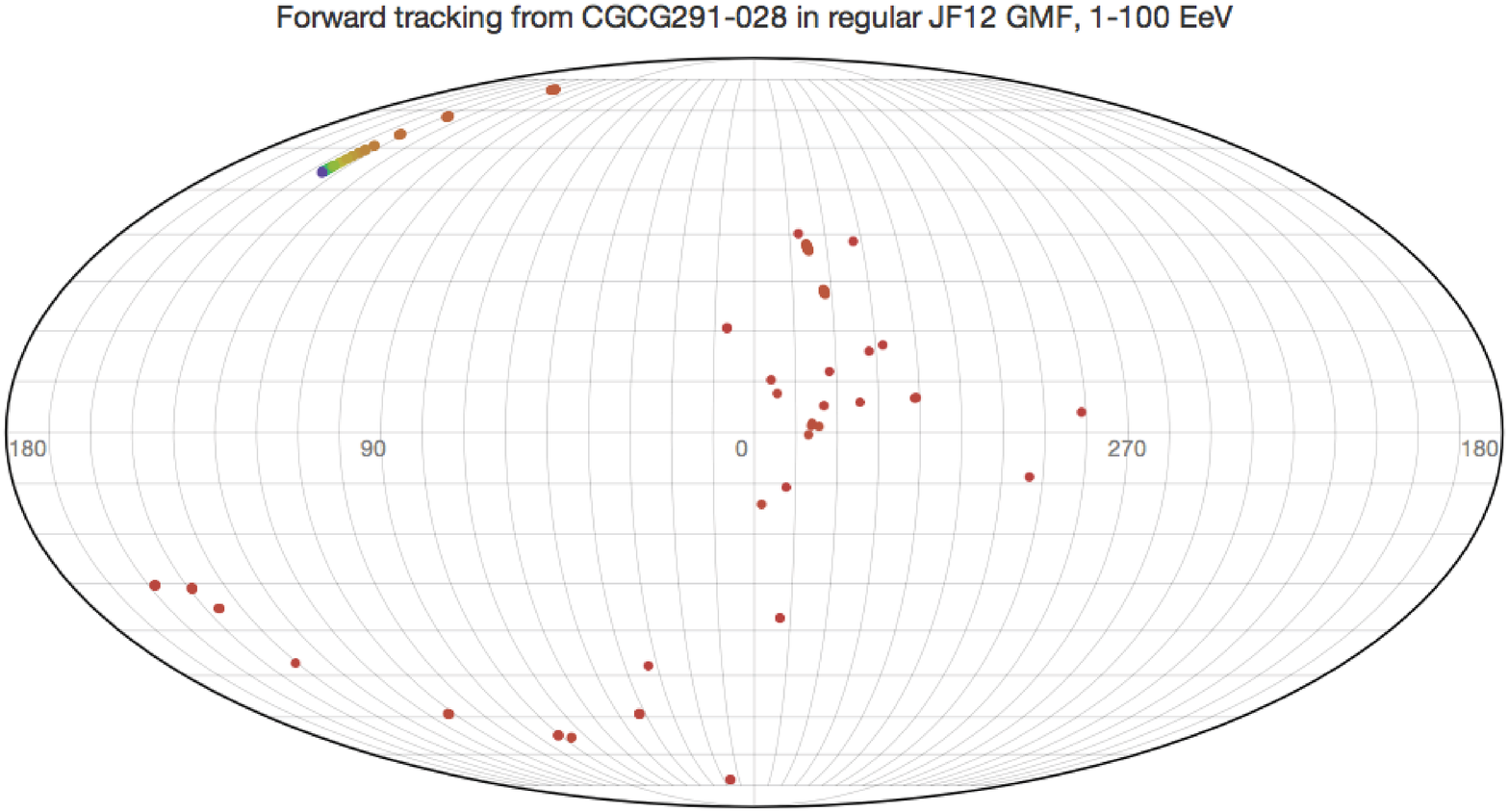}
\includegraphics[width=0.4\textwidth]{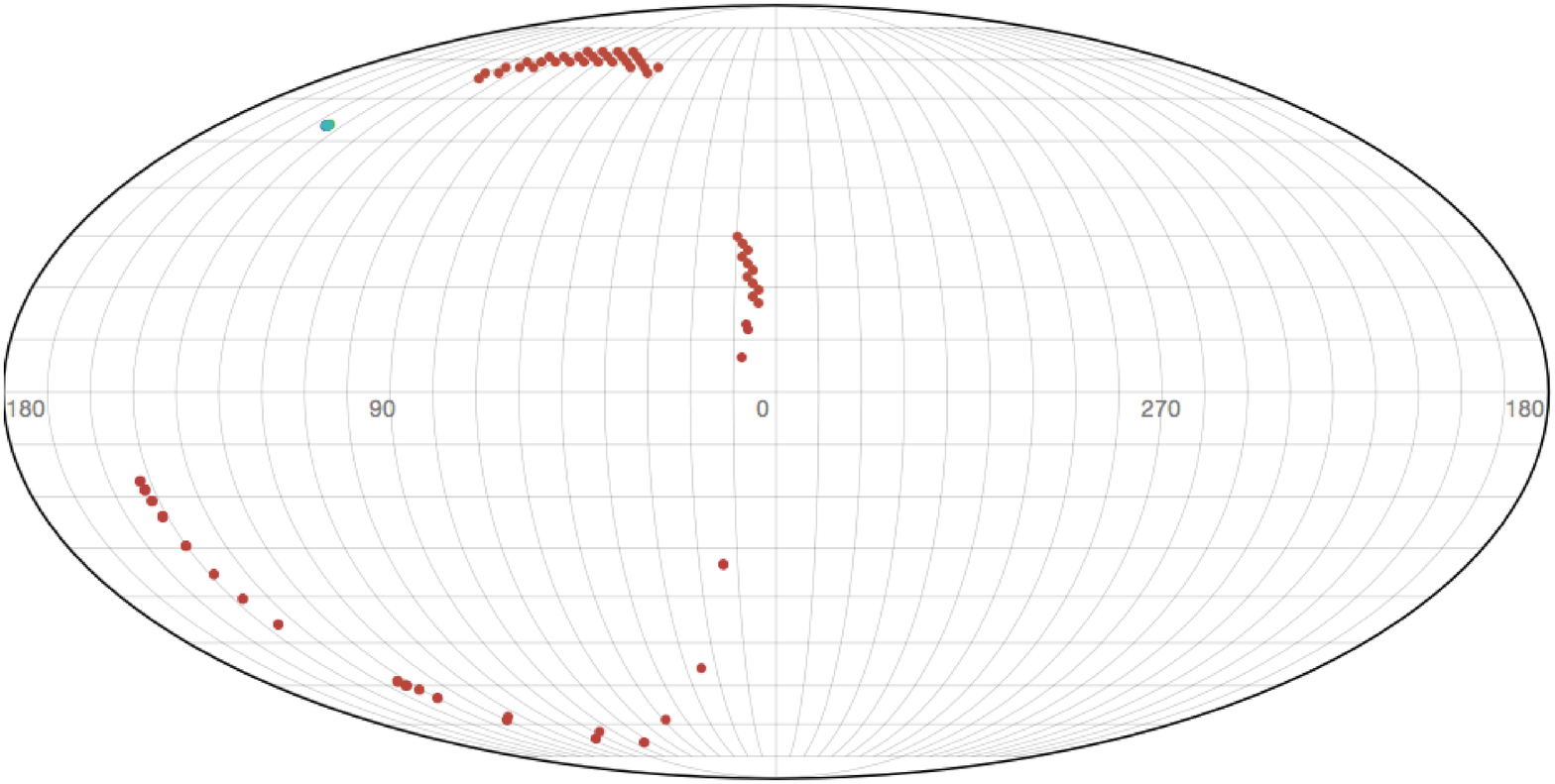}
\caption{\label{cgcgdefs} Locus of UHECRs from \cgcg\  in Galactic coordinates due to deflections in the coherent JF12  GMF, for rigidity values from 100 EV to 1 EV in steps of log$(R$)=0.05 (above), and in the JF12 field with $B_{X}\equiv 0$ for a subset of $R$ values down to 1.4 EV.}
\end{figure}

Henceforth we will assume the four UMC events are protons from a common source.  The star in Fig. \ref{uhecr_image} shows the position to which they converge when backtracked in the GMF.  Fig. \ref {cgcgdefs} shows the predicted locus of events coming from this direction for rigidities from 100 to 1 EV in steps of 0.05 in log$(R$), deflected in the JF12 coherent field (upper panel).  For rigidities below about 20 EV the deflections become very large and display a large dispersion even without random fields, due to multiple imaging.  Including the random field will increase the variance, but for rigidity $\geq 35$ EV, the dispersion from the random field can be expected to be of order a degree or so;  quantifying this will be addressed in future work.  It is interesting that the ``stream'' of events at lower rigidities falls qualitatively along the direction of the supergalactic plane so the same source could  conceivably contribute to the excess of events in the Cen A region observed by Auger \cite{AugerUpdate} and the excess recently reported by TA correlated with the supergalactic plane\cite{TinyakovICRC13}.   To give an idea of the sensitivity to uncertainties in the coherent GMF, the deflections with the ``$X$-''field set to zero keeping other field parameters fixed, are shown in the lower panel of  Fig. \ref {cgcgdefs} for a sampling of log($R)$ values; for $Z=1$ the UMC deflections are similar to those for the JF12 field, as argued above, and even for low rigidities the qualitative shape of the stream is similar albeit with less multiple imaging.

\vspace{-0.2 in}
\section{Tentative identification of the source of the Ursa Major UHECRs} \label{src}

In order to confine CRs during the acceleration process, their Larmor radius must remain smaller than the size $R_{acc}$ of the accelerating system, placing a lower bound on $B \times R_{acc}$.  This gives a lower bound on the Poynting luminosity in the accelerator which implies a lower bound on the bolometric accretion luminosity accompanying UHECR acceleration\cite{fg09} if the accelerator is powered by accretion.   The minimum bolometric luminosity for perfect efficiency is $\approx 10^{{45}} (E/{Z\times100 \rm EeV})^{2}$ erg/s  -- achievable in GRBs and the most powerful AGNs.  However sufficiently powerful AGNs and GRBs are too rare to account for the UHECR flux and observationally-inferred bounds on the density of sources if a significant fraction of UHECRs are protons\cite{fg09}, as suggested by the correlation with (weak) AGNs found by Auger\cite{augerScience}.  This conundrum prompted the proposal that a significant fraction of UHECRs are produced in a new class of powerful transients \cite{fg09}, as could arise from the tidal disruption of a star by a supermassive black hole.  Examples of such flares have subsequently been discovered\cite{vf11}, particularly noteworthy being \swift\  J164449.3+573451\cite{SwiftTDF} in blazar mode. 

There is no powerful enough AGN within the $z \approx 0.07$ GZK horizon and $< 5^{\circ}$ of the UMC source direction; since we have deduced that the four Ursa Major events are protons, it follows that they were accelerated in a transient event because as we shall see below, the delay between light and UMC events' arrival is too small for a powerful AGN to have faded in that interval.  If the source was a GRB we would not expect any visible relic, but if the source was a tidal disruption or other new type of powerful but transient flare, there might be nuclear activity observable today.  After a tidal disruption flare (TDF) the residual activity may subside slowly enough to remain visible, or the TDF rate may be enhanced in AGNs (even those with low-luminosities), so that UHECR sources may preferentially have low-level AGN activity.    

The above reasoning, which suggests that the host of the UMC events may presently be a low-luminosity AGN, along with the correlation of Auger UHECRs with \swift-BAT AGNs, whose luminosities are typically $\lesssim 10^{43}$ erg/s, motivated the author to look for a BAT AGN with $z < 0.07$ within the $1^{\circ}$ locus of the UMC source.  There was no such source listed in the existing (58 month) \swift-BAT AGN catalog, but the hard X-ray source \swift\ J$1105.7+5854$, which the \swift-BAT team had associated with a galaxy at $z=0.192$ (far beyond the GZK horizon) is in the angular domain of interest.  The author checked for an alternative host candidate at lower redshift and found that the galaxy CGCG-291-028, at $z=0.0497$, is within the error radius of \swift\ J$1105.7+5854$.  An inquiry determined that the 70 month \swift-BAT AGN catalog (subsequently made public \cite{BAT70mo}) resolves two sources, one of which was near but possibly not inside CGCG 291-028.  A recent \chandra\ observation obtained by GRF and W. Baumgartner has shown that the X-ray souce is unambiguously within CGCG 291-028, and very close to or coincident with its nucleus.  Thus CGCG 291-028 is the unique and compelling candidate for being the source of the UMC events.  Additional observations of CGCG 291-028 at optical and radio wavelengths have also been obtained and will be described elsewhere.

\vspace{-0.2 in}
\section{Transient source}\label{burst}
Due to deflections in the random extragalactic magnetic field, a UHECR of these energies arrives centuries or more after simultaneously emitted photons, so there is no possibility of corroborating the transient-acceleration scenario through seeing an electromagnetic counterpart of a UHECR burst.  However the spectrum of UHECRs from a single source carries crucial information on the nature of the acceleration process.  The observed spectrum from an individual continuous source such as a powerful AGN is approximately the same as the all-sky UHECR spectrum, $\sim E^{{-2.7}}$, whereas a bursting source has a sharply peaked spectrum with an energy spread of only a factor of about 2 \cite{waxmanME}.  This is because a UHECR's magnetic deflections and hence delay time decrease with energy.  Therefore those UHECRs arriving in a time interval $<<\Delta T$, a time $\Delta T$ after photons from the burst, have a relatively similar energy: the higher energy CRs will have already passed and lower energy CRs will not yet have arrived.   

\begin{figure}[t!]
\centering
\includegraphics[width=0.4\textwidth]{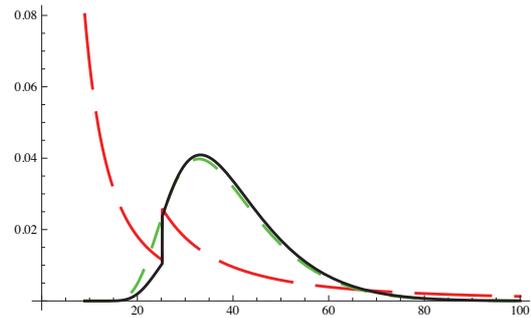}
\caption{\label{spec} Shape of observed power law (red, long dash) and transient (black, solid) spectra given the AGASA+HiRes exposure; the dashed green line shows the transient spectrum for a uniform exposure.}
\end{figure}

Thus two independent pieces of evidence point to the UMC events having been produced in a burst or flare:  \\
$\bullet$ The event energies fall in a narrow range, 35-50 EeV, whereas if the spectrum were a simple power law, $dN/dE \sim E^{-p}$, there should be 1.5 times more events in say the 10-25 EeV range than above 25 EeV, for $p = 2.7$ and uniform exposure.   Fig. \ref{spec} shows the expected spectrum in the transient and continuous cases, given the energy-dependent HiRes and AGASA exposures\cite{f07}.  Accounting for the energy dependent exposure, 5.5 are expected between 10-25 EeV when 4 are seen above, if the source were continuous; the probability of finding 0 events when 5.5 are expected for a continous source is 0.4\%, whereas the event energies (Table \ref{UMevts}) are are completely natural for a transient spectrum (Fig. \ref{spec}).   Below 10 EeV the data is not public, but in any case for rigidities below 10 EV there start to be multiple images so the locus of events from the UMC source cannot be determined with sufficient confidence to extend the search to lower rigidity. Random fields should be included along with the regular GMF, and better techniques are needed for identifying streams in the much larger sample of lower energy events, most of which are background from distant sources.  
\\
$\bullet$  The bolometric luminosity of  \cgcg\  today, $\approx 10^{44}$erg/s based on its 2-10 keV luminosity recently measured with \chandra\ (W. Baumgartner and GRF, in preparation), falls significantly short of the minimum required to accelerate a proton to the observed energies assuming perfect efficiency, $
\Gamma^{2} \,\, 3 \times 10^{44}$erg/s\cite{fg09}, where 
$\Gamma \gsi 1$ is the bulk Lorentz factor of shocks in the jets.  A similar luminosity shortfall pertains in almost all of the AGNs which have been found to correlate with Auger UHECRs\cite{zfg09,tzf12}.  \\

\vspace{-0.2 in}
\section{Implications}\label{interp}

The light propagation time from the UMC is $6 \times 10^{8}$ yr, and the UHECR arrival time delay relative to this is
\be
\Delta T \sim D/c \, (1-cos(\theta_s)^2) \sim D \theta_s^2/(2 c) 
\label{DeltT}
\ee
where
\be
\label{thetas}
\theta_s(E) =  \frac{\sqrt {2 D q^2 \langle B^2 \lambda \rangle}}{3 E}.
\ee
Here $\lambda$ is the characteristic maximum coherence length of the extragalactic magnetic turbulence and $B$ is its RMS strength.  
If the UHECRs in the UMC were produced by \cgcg\ when it was in a flaring state, we can draw some powerful conclusions.  Qualitatively, the flux of events observed at a distance $D$ from a bursting source is
\begin{eqnarray*}
\label{qualitative}
F_{\rm obs} &  \sim  \frac{{\rm (Number \,of \,events \,produced \,at \,the \,source})}{4 \pi D^2 \,{\rm (Spread \,in \,arrival \,times)}} \\
  & \sim \frac{N(E)}{4 \pi D^2 D \theta_s^2/(2 c)} \sim \frac{9 c q N(E) E^2}{4 \pi D^3 D  q^2 \langle B^2 \lambda \rangle },
\end{eqnarray*}
where we have used eqns (\ref{DeltT} ) and (\ref{thetas}).  In the limit of many small deflections, the problem has been solved analytically in \cite{alcockH,waxmanME} and the quantitative expression for the flux when $Z=1$ is
\be
\label{flux}
F(E; D, E_0) =  \frac{3 c E^2 N(E) }{8 \pi  \langle B^2 \lambda \rangle D^4} G_{A\rm H}( (E/E_0)^2),
\ee
where $G_{\rm AH}$ is the normalized probability distribution function given in \cite{alcockH}.  The parameter $E_0$ is determined by the distance and time delay of the observation, and the magnetic structure of the intervening medium:  
\be
\label{E0}
E_0 \equiv D  \left( \frac{2 q^2 \langle B^2 \lambda \rangle}{3 c \Delta T} \right)^{1/2}.
\ee
The peak of the spectrum is at $E_{\rm peak} =  0.214\, E_0$ and the average energy is $\bar{E} =\,  0.25 \,  E_0$.  From the latter, we infer that for the UMC today, $E_{0} = 145$ EeV.  Since we know the distance to \cgcg\, $D \approx 200$ Mpc, if we knew $\langle B^2 \lambda \rangle$ we could infer $\Delta T$ or vice versa.

In principle, we could determine $\langle B^2 \lambda \rangle$ if the extragalactic magnetic smearing could be measured, since we know the distance and could use (\ref{thetas}).  However $\theta_s(E)$ may well be comparable to or smaller than the angular dispersion from the random Galactic magnetic field and present observational resolution, so even with more events and better angular resolution, at best we could place an upper limit on $\theta_s(E)$.  

However we have additional information we can exploit.  The total energy in the UMC protons observed by AGASA and HiRes is 160 EeV.  Equating this to the integral of the exposure-weighted energy flux (\ref{flux}) gives the total energy at the source in UHECR protons above $E_{\rm min} = 1$ EeV:
\be
E_{{\rm src}} = 9 \times 10^{56} {\rm erg} \langle B_{\rm nG}^2 \lambda_{\rm Mpc} \rangle,
\ee
for a spectral index $E^{-2.3}$.  Since the rest-mass energy of a 1 $M_{\odot}$ is $2 \times 10^{54}$ ergs and the maximum mass of a star which is reasonably likely to be disrupted is $\sim 10 M_{\odot}$, and the fraction of that energy going into UHE protons above 1 EeV is unlikely to be greater than 10\%, we derive an upper bound on the extragalactic magnetic field:
\be
B_{\rm EG, rms} < 1.4 \times 10^{-11} \lambda_{\rm Mpc}^{-1/2}.
\ee
With this, we can estimate the extragalactic smearing angle to be $< 0.15^{\circ}$ and place an upper bound on the time-delay since the arrival of the photons of the flare of $\Delta T < 2000  {\rm yr}$.

\vspace{-0.2 in}
\section{Summary}

We have shown that the Ursa Major cluster of 4 events in the combined AGASA and HiRes datasets, when backtracked in the Galactic magnetic field under the assumption that they are protons, become even more tightly clustered and have an rms separation of 0.8$^{\circ}$ with respect to the galaxy CGCG 291-028 at redshift $0.0471$ (distance of $\approx 200$ Mpc), which is the host of a hard X-ray AGN.  The spectrum of the events are consistent with production in a transient, while the absence of events in the 10-35 EeV range would be difficult to explain if their source were dominantly continuous.  The total power in UHECRs inferred from the observed flux is compatible with what is available in a stellar tidal disruption.  Combined with the recently inferred rate\cite{vf12} and luminosity\cite{f12} of tidal disruption events, the UMC data fit into a coherent picture in which tidal disruption events produce a significant portion of UHECRs\cite{fg09,f12}.

The UMC events were found with a total exposure of 2850 km$^{2}$-yr, so if the UMC cluster is not a statistical fluke, Telescope Array has a good chance of finding events in the same region and energy range.  It would of be of interest to search for lower rigidity events from the same source as would arise from an earlier flaring episode or continuous acceleration of high-$Z$ nuclei in a lower luminosity steady-state.  Unfortunately, even if a lower rigidity stream exists it would likely be quite broad and its position cannot presently be predicted with adequate accuracy.  It is intriguing however that the stream predicted wtih the JF12 field alone (Fig. \ref{cgcgdefs}) is in roughly the right location to contribute to excesses reported by Auger and TA, if there were a heavy component to the spectrum at the source (Sec. \ref{magdefs}).

This work was supported in part by NSF PHY-1212538; I wish to especially thank W. Baumgartner and J. Roberts for their contributions.


\end{document}